\documentclass[
    ,final            
  ,numberedheadings 
  ]
  {aipproc}

\usepackage{makeidx}
\usepackage{theorem,amsmath,amssymb,mathrsfs}
\usepackage{float,supertabular}

\usepackage{mathrsfs}

\layoutstyle{8x11single}

\def\index#1{}
\def\d{\operatorname{d}}\def\<{\langle}\def\>{\rangle}
\def\:{\hbox{\bf :}}
\def\Reals{\mathbb R}

\def\set#1{{\sf #1}}

\def\rnk{\operatorname{rank}}
\def\Op#1{\operatorname{Op}_{#1}}
\def\Conv{\set{Co}}

\def\dim{\operatorname{dim}}\def\idim#1{\operatorname{idim}(#1)}
\def\cdim#1{\operatorname{cdim}(#1)}\def\mdim#1{\operatorname{mdim}(#1)}
\def\Extr{\operatorname{Extr}}\def\chao#1{\chi(#1)}
\def\face#1{\operatorname{Fc}(#1)}

\def\sH{\set{H}}

\def\qed{$\,\blacksquare$\par}
\def\eg{e. g. }\def\ie{i. e. }
\def\n#1{|\!|#1|\!|}

\newtheorem{definition}{Definition}
\newtheorem{lemma}{Lemma}
\newtheorem{conjecture}{Conjecture}

\newtheorem{corollary}{Corollary}
\newtheorem{theorem}{Theorem}
\newtheorem{gaxiom}{General axiom}
\newtheorem{grule}{Rule}
\def\Proof{\medskip\par\noindent{\bf Proof. }}
{\theorembodyfont{\rmfamily}{\newtheorem{remark}{\bf Remark}}}
{\theorembodyfont{\rmfamily}{\newtheorem{example}{\bf Example}}}
\def\Quote#1#2{
\begin{itemize}
\item[] {\em #1}
\item[] \hfill ---{ #2}
\end{itemize}}
\def\trnsfrm#1{\mathscr #1}
\def\tA{\trnsfrm A}\def\tB{\trnsfrm B}\def\tC{\trnsfrm C}\def\tS{\trnsfrm S}
\def\tI{\trnsfrm I}\def\tT{\trnsfrm T}\def\tU{\trnsfrm U}

\def\AA{\mathbb A}\def\AB{\mathbb B}\def\AC{\mathbb C}\def\AL{\mathbb L}
\def\cA{{\underline{\tA}}}\def\cB{\underline{\tB}}

\def\Stset{{\mathfrak S}}\def\Trnset{{\mathfrak T}}\def\Actset{{\mathfrak A}}
\def\Cntset{{\mathfrak P}}\def\Prdset{{\mathfrak P}_p}
\def\glossaryentry#1#2{#1 & #2 \\}\def\Idx#1{%
}

\makeglossary
\begin{document}

\title{On the missing axiom of Quantum Mechanics}


\author{Giacomo Mauro D'Ariano}{
  address={{\em QUIT} Group, Dipartimento di Fisica ``A. Volta'', via Bassi 6,
I-27100 Pavia, Italy, {\em http://www.qubit.it}\\
 Department of Electrical and Computer Engineering, Northwestern University, Evanston,   IL  60208}}

\begin{abstract}
The debate on the nature of quantum probabilities in relation to Quantum Non
  Locality has elevated Quantum Mechanics to the level of an {\em Operational Epistemic Theory}. In
  such context the quantum superposition principle has an extraneous non epistemic nature. This leads
  us to seek purely operational foundations for Quantum Mechanics, from which to derive the
  current mathematical axiomatization based on Hilbert spaces. 
\par In the present work I present a set of axioms of purely operational nature, based on a general 
definition of "the experiment", the operational/epistemic archetype of information retrieval from
reality. As we will see, this starting point logically entails a series of notions [state,
conditional state, local state, pure state, faithful state, instrument, propensity 
(i.e. "effect"), dynamical and informational equivalence, dynamical and informational compatibility,
predictability, discriminability, programmability, locality, a-causality, rank of the state, 
maximally chaotic state, maximally entangled state, informationally complete propensity, etc. ],
along with a set of rules (addition, convex combination, partial orderings, ... ), which, far from
being of quantum origin as often considered, instead constitute the universal {\em syntactic
  manual} of the operational/epistemic approach. The missing ingredient is, of course, the quantum
superposition axiom for probability amplitudes: for this I propose some substitute candidates of
purely operational/epistemic nature. 
\end{abstract}
\maketitle
\section{Introduction}
Quantum Mechanics is not as any other physical theory. It applies to the entire physical domain,
from micro to macro-physics, independently of the size and the energy scale, from particle physics, to nuclear,
atomic, molecular, solid state physics, from the tiniest particle, to cosmology. Despite such
generality, Quantum Mechanics still lacks a physical axiomatization---a quite embarrassing situation
when we teach the theory to students.  Why so abstract mathematical objects such as ``Hilbert
spaces'' stay at the core axiomatic level of our most general physical theory? We are used to answer:
``This is the {\em quantum superposition principle}, which entails complementarity and wave-particle
dualism''. That way we save our face.  

In its very essence Quantum Mechanics addresses, for the first time, the core problem of Physics:
that of {\em Measurement}. More generally, I would say, Quantum Mechanics deals with the description
of the {\em Physical Experiment}. The probabilistic framework, which, in such context, is generally
dictated by the obvious need of working in the presence of incomplete knowledge, contrarily to our
original intentions turns out to be not of {\em epistemic} nature, but is truly {\em ontic}. This is
the lesson of nonlocal EPR correlations. Incredibly, ``God actually plays dice!'' Now, this makes
the situation even more embarrassing: on the basis of the quantum superposition principle of
probability amplitudes we ``physicists'' preach the ontic nature of probability, and elevate
Quantum Mechanics to a ``Theory of Knowledge''! 

Clearly, in this new view, the quantum superposition principle is not an acceptable starting point
anymore: for a Theory of Knowledge we should seek operational axioms of epistemic nature,
and be able to derive the usual mathematical axiomatization from such operational axioms. Shortly:
for a Theory of Knowledge we need Axioms of Knowledge.

In the present work my starting point for this axiomatization is the definition of {\em what an
  experiment is}. Indeed, ``the experiment'' is the archetype of the {\em cognitive act}, being the
prototype interaction with reality able to get information on it. As we will see, adopting a general
definition of experiment that includes all possible interactions and information exchanges with
reality, is a very seminal starting point, which logically entails a series of notions---such as
that of state, conditional state, local state, pure state, faithful state, instrument, propensity
(i.e. "effect"), dynamical and informational equivalence, dynamical and informational compatibility,
predictability, discriminability, programmability, locality, a-causality, rank of a state,
etc. ]---along with a set of rules (addition, convex combination, partial orderings, ... ), which,
far from being of quantum  origin as often considered, instead constitute the universal {\em
 syntactic manual} of the cognitive/operational approach. The missing ingredient is still, of
course, the quantum superposition axiom, and for this I will propose at the end some substitute
candidates of purely cognitive/operational nature. 

In the present attempt some expert readers will recognize similarities with the program of other authors
during the seventies, following the Ludwig school \cite{Ludwig-axI}, which were seeking operational
principles to select the structure of quantum states from all possible convex structures [see, for
example, the papers of  U. Krause \cite{krause74}, H. Neumann \cite{Neumann74}, and E. St\o rmer
\cite{Stormer74} collected in the book \cite{hartkamper74}]. Why these work didn't have a followup?
I think that, besides the fact that the convex structure by itself is not sufficiently rich mathematically for 
deriving an underlying Hilbert space structure, concepts as {\em entanglement} and {\em
informationally complete measurements} (i. e. quantum tomography \cite{tomo_lecture}) were still not
familiar in those days. Recently it has been shown that it is possible to make a complete quantum calibration of a
measuring apparatus \cite{calib} or of a quantum operation \cite{tomo_channel} by using a single pure
bipartite state. I think that this gives us a new unique opportunity for deriving the Hilbert space structure
from the convex structure in terms of {\em calibrability axioms}, which relies on the special link
between the convex set of transformations and that of states which occurs in Quantum Mechanics,
and which make the transformations of a single system resemble closely states of a bipartite
system \cite{choi75,Jamiolkowski72} 
\section{Axioms for the experiment}
\Quote{It is the theory which decides what we can observe!}{Einstein to Heisenberg}
\medskip
\begin{gaxiom}[On inductive-deductive science]\label{ga:1}
  In any experimental inductive-deductive science we make {\em experiments} to get {\em information}
  on the {\em state} of a {\em objectified physical system}. Knowledge of such a state will allow us to
  predict the results of forthcoming experiments on the same object system. Since we necessarily work
  with only partial {\em a priori} knowledge of both system and experimental apparatus, the rules
  for the experiment must be given in a probabilistic setting.
\end{gaxiom}
Notice that the information is of the {\em state} of the system, not of the system itself. In fact,
in order to set the experiment we need some prior information on the physical system, e. g. if it is
an electric current, a field, or a particle, what is its charge, etc. The goal of the experiment is
to determine something unknown (or imprecisely known) about the system: logically this should
enter in the notion of state, as will be given in Def. \ref{istate}. The boundary between
what is the object and what is its state will depend on the context of the particular experiment,
e. g. the charge of a particle can be a property defining the object system---and used to design the
measuring apparatus; if unknown, a property could be the object of the experiment itself, and, as such, it
would enter the definition of state. Again we emphasize that our purpose is to give only the
syntactic manual of the empirical approach, not the semantics, i. e.  the specific physical context.
\bigskip
\begin{gaxiom}[On what is an experiment]\label{ga:2} An experiment on an
  object system consists in having it interact with an apparatus. The interaction between object
  and apparatus produces one of a set of possible transformations of the object, each one occurring
  with some probability. Information on the ``state'' of the object system at the beginning of the
  experiment is gained from the knowledge of which transformation occurred, which is the "outcome"
  of the experiment signaled by the apparatus.
\end{gaxiom}
It is clear that here "object" and "apparatus" are both physical systems, and the asymmetry between
object and apparatus is just an asymmetry in prior knowledge, namely the apparatus is the system of
which the experimenter has more prior information. Clearly the knowledge gained about the state of the
object depends also on the knowledge of details of the transformation undergone by the object system,
and, generally, also on preexisting knowledge of the system ``state'' itself. In other words,
the experiment can be always regarded as a {\em refinement} of knowledge on the object system.
\medskip\par One should convince himself that the above definition of
experiment is very general, and includes all possible situations. For example, at first sight it may
seem that it doesn't include the case in which the object is not under the experimenter's control
(\eg astronomical observations), in the sense that in such case one cannot establish an interaction
with the object system. However, here also there is an interaction between the object of
interest (\eg the astronomical object) and another object (\eg the light) which should be regarded
as a part of the apparatus (\ie telescope$+$light). Such cases can also be regarded as "indirect
experiments", namely the experiment is performed on an auxiliary "object" (\eg the light) which is
supposed to have experienced a previous interaction with the ultimate object of interest, and whose state
depends on properties/quantities of it.  Also, the customary case in which a "quantity" or a
"quality" is measured without in any way affecting the system corresponds to the case in which all
states are left invariant by the transformations corresponding to each outcome.  \bigskip
\par Performing a different experiment on the same object obviously
corresponds to the use of a different experimental apparatus or, at least, to a change of some settings of the
apparatus. Abstractly, this corresponds to change the set $\{\tA_j\}$ of possible transformations,
$\tA_j$, that the system can undergo.
Such change could actually mean
really changing the "dynamics" of the transformations, but it may simply mean changing only their
probabilities, or, just their labeling outcomes. Any such change actually corresponds to a change of
the experimental setup. Therefore, the set of all possible transformations $\{\tA_j\}$ will be
identified with the choice of experimental setting, i.~e. with the {\em experiment}
itself---or, equivalently, with the {\em action} of the experimenter: this will be formalized by the
following definition 
\begin{definition}[Actions/experiments and outcomes]\glossary{\Idx{actions}$\AA,\AB,\AC,\ldots$ & actions}
\index{action!definition}\index{outcome} An {\bf action} or {\bf experiment} on the
object system is given by the set $\AA\equiv\{\tA_j\}$ 
of possible transformations $\tA_j$ having overall unit probability,
with the apparatus signaling the {\bf outcome} $j$ labeling which
transformation actually occurred.  
\end{definition}
Thus the action/experiment is just a {\em complete} set of possible transformations that can
occur in an experiment.
As we can see now, in a general probabilistic framework the {\em action} $\AA$ is the "cause",
whereas the {\em outcome} $j$ labeling the transformation\index{cause and effect} $\tA_j$ that
actually occurred is the "effect". The {\em action} has to be regarded as the ``cause'', since it is the
option of the experimenter, and, as such, it should be viewed as deterministic (at least one
transformation $\tA_j\in\AA$  will occur with certainty), whereas the outcome $j$---\ie which
transformation $\tA_j$ occurs---is probabilistic.  
The special case of a deterministic transformation $\tA$ corresponds to a {\em singleton
  action/experiment} $\AA\equiv\{\tA\}$. 
\index{transformation!deterministic}
\medskip
\par In the following, wherever we consider a nondeterministic transformation $\tA$ by itself, we always regard it in
the context of an experiment, namely for any nondeterministic transformation there always exists a
at least complementary one $\tB$ such that $\omega(\tA)+\omega(\tB)=1$ for all states $\omega$.
\section{States}
According to General Axiom \ref{ga:1} by definition the knowledge of the state of a physical system
allows us to predict the results of forthcoming possible experiments on the system, or, more
generally, on another system in the same physical situation. Then, according to the General Axiom
\ref{ga:2} a precise knowledge of the state of a system would allow us to evaluate the probabilities of
any possible transformation for any possible experiment. It follows that the only possible
definition of state is the following
\begin{definition}[States]\label{istate}\index{state(s)}
\glossary{\Idx{state1}$\omega,\zeta,\ldots$ & states}
\glossary{\Idx{state2}$\Omega,\Phi,\ldots$ & multipartite states}
 A  state $\omega$ for a physical
  system is a rule that provides the probability for any possible
  transformation, namely 
\begin{equation}
\omega:\textbf{state},\quad\omega(\tA):\text{probability that the
  transformation $\tA$ occurs}.
\end{equation}
\end{definition}
\medskip
We assume that the identical transformation $\tI$ occurs with probability one, namely
\glossary{\Idx{transformations2}$\tI$ & identical transformation}
\begin{equation}
\omega(\tI)=1.\label{normcond}
\end{equation}
This corresponds to a kind of {\em interaction picture}, in which we don't consider the free
evolution of the system\index{intermediate/interaction picture}
(the scheme could be easily generalized to include a free evolution). Mathematically, a state will
be a map $\omega$ from the set of physical transformations to the
interval $[0,1]$, with the normalization condition
(\ref{normcond}). Moreover, for every action $\AA=\{\tA_j\}$ 
one has the normalization of probabilities \index{action!normalization condition}
\begin{equation}
\sum_{\tA_j\in\AA}\omega(\tA_j)=1
\end{equation}
for all states $\omega$ of the system. As already noticed, in order to
include also non-disturbing experiments, one must conceive 
situations in which all states are left invariant by each
transformation (see also Remark \ref{r:noinfo} in the following).
\medskip
\par The fact that we necessarily work in the presence of partial knowledge about both object and
apparatus requires that the specification of the state and of the transformation could be given
incompletely/probabilistically, entailing a convex structure on states and an addition rule for
coexistent transformations.  The convex structure of states is given more precisely by the rule
\begin{grule}[Convex structure of states]\label{idim}\index{state(s)!convex structure}
The possible states of a physical system comprise a convex set: for any two states 
$\omega_1$ and $\omega_2$ we can consider the state $\omega$ which is
the {\em mixture} of $\omega_1$ and $\omega_2$, 
corresponding to have $\omega_1$ with probability $\lambda$ and
$\omega_2$ with probability $1-\lambda$. We will write
\begin{equation}
\omega=\lambda\omega_1+(1-\lambda)\omega_2,\quad 0\le\lambda\le 1,
\end{equation}
and the state $\omega$ will correspond to the following probability
rule for transformations $\tA$
\begin{equation}
\omega(\tA)=\lambda\omega_1(\tA)+(1-\lambda)\omega_2(\tA).
\end{equation}
\end{grule}
Generalization to more than two states is obtained by induction. In the
following the convex set of states will be denoted by $\Stset$. 
\glossary{\Idx{convex1}$\Stset$ & convex set of states}
We will call {\em pure} the states which are the extremal elements of
the convex set, namely which cannot be obtained as mixture of any two
states, and we will call {\em mixed} the non-extremal ones. As regards
transformations, the addition of coexistent transformations and
the convex structure will be considered in Rules \ref{g:addtrans} and \ref{r:convextrans}.
\bigskip
\par Recall that for the convex set of states, as for any convex
set, one can define partial orderings as follows.
\begin{definition}[Partial ordering of states]\label{d:part-ord}\index{state(s)!partial ordering of} For
  $\omega,\zeta\in\Stset$, $\alpha\in[0,1]$, denote by 
\begin{enumerate}
\item $\omega\prec_\alpha \zeta$ if there exists a $\theta\in\Stset$ such that $\zeta=\alpha \omega+(1-\alpha)\theta$;
\item $\omega\sim_\alpha \zeta$ if $\omega\prec_\alpha \zeta$ and $\zeta\prec_\alpha \omega$;
\item $\omega\prec \zeta$ if there exists $\alpha>0$ such that $\omega\prec_\alpha \zeta$;
\item $\omega\sim \zeta$ if $\omega\prec \zeta$ and $\zeta\prec \omega$.
\end{enumerate}
\glossary{\Idx{prec1}$\prec_\alpha$ & partial ordering of states}
\glossary{\Idx{prec2}$\prec$ & partial ordering of states}
\end{definition}
For example, we can "read" the definition of $\prec$ in the following
way: $\omega\prec \zeta$ means that $\omega$ belongs to an ensemble 
for $\zeta$.
\begin{definition}[Minimal decomposition of a state] 
\index{minimal convex decomposition}\index{convex decomposition!of state, minimal}
A minimal convex decomposition of a state is a convex expansion of the state in a minimal set of extremal states.
\end{definition}
\begin{definition}[Caratheodory rank of a state]
\index{state(s)!rank}\index{rank of a state}
\glossary{\Idx{dimension1}$\rnk(\omega)$ & Caratheodory rank of the state $\omega$}
 The Caratheodory rank $\rnk(\omega)$ of the state $\omega\in\Stset$ (or simply rank) is the
minimum number of extremal states in terms of which we can write the state as convex
combination. This is also given by $\dim[\face{\omega}]+1$, where 
$\face{\omega}\subseteq\partial\Stset $ is the "face" to which the state $\omega$ belongs. 
\glossary{\Idx{convex7}$\face{\omega}$ & face of the convex set of states to which $\omega$ belongs}
\glossary{\Idx{convex8}$\partial\Stset$ & border of $\Stset$}
\end{definition}
\begin{definition}[Caratheodory dimension]
\index{Caratheodory dimension}\index{dimension!Caratheodory}
\glossary{\Idx{dimension2}$\cdim{\Stset}$ & Caratheodory dimension of the convex set of states $\Stset$}
 We call the maximal rank of a state in $\Stset$ the {\em Caratheodory dimension} of $\Stset$, denoted by
 $\cdim{\Stset}$. 
\end{definition}
\begin{remark}\index{Caratheodory's theorem}
  According to the Caratheodory's theorem, for a convex set of real affine dimension $n$ (i. e.
  embedded in $\Reals^n$) one needs at most $n+1$ extremal points to specify any point of the convex
  set as convex combination. However, for the convex sets of Quantum Mechanics one needs much fewer
  extremal points, precisely only $\sqrt{\dim(\Stset)+1}$ (the convex sets of states in Quantum
  Mechanics have real affine dimension $\dim(\Stset)=k^2-1$, $k$ being the dimension of the Hilbert
  space). Therefore, only 
  $\sqrt{\dim(\Stset)+1}$ pure states are necessary to express each state as a convex combination.
  Such states are also a maximal set of perfectly discriminable states (see the following).
\glossary{\Idx{dimension5}$\dim(\Stset)$ & affine dimension of the convex set of states  $\Stset$}
\end{remark}
\begin{remark}
It is worth noticing that the dimension of the faces of the full convex set of quantum states $\Stset$ 
for given finite dimension of the underlying Hilbert space decreases discontinuously
in quadratic ladders. For example, the 8 dimensional convex set of states (corresponding to Hilbert
space dimension $d=3$) has faces that are 3-d Bloch spheres. Therefore, the faces of a complete set
of quantum states are themselves complete set of quantum states (for lower dimension of the
underlying Hilbert space). Each face of the complete convex set of states is itself a complete
convex set of states at lower Hilbert space dimension. This lead us to consider also the following rule
\end{remark}
\begin{grule}
The faces of a "complete" set of states are themselves "complete" sets of states.
\end{grule}
The above rule needs a definition of what we mean by "completeness", and a possible route could be
via the action of all possible invertible dynamical maps, i. e. the isometric indecomposable
transformations of the set of states, namely the equivalent of unitary transformations (see the
following). Notice, however, that the notion of {\em completeness} is not strictly operational, and
for this reason we will not pursue this axiomatic route.
\par\medskip Using the partial ordering on the convex set of states we can easily define the
maximally chaotic state as follows
\begin{definition}[Maximally chaotic state]\label{def:chaostate} The maximally chaotic state
  $\chao{\Stset}$ of $\Stset$ is the most mixed state of $\Stset$, in the sense that
\begin{equation}
\forall\theta\in\Stset\qquad \max\{\alpha\in[0,1]\,:\theta\succ_\alpha\chao{\Stset}\}\ge
 \max\{\beta\in[0,1]\,:\chao{\Stset}\succ_\beta\theta\}.
\end{equation}
\end{definition}
An alternative definition is that of baricenter-state
\begin{definition}[Alternative definition of maximally chaotic state]\label{def:chaostate2}
\glossary{\Idx{chao}$\chao{\Stset}$ & maximally chaotic state of the convex set of states $\Stset$}
\index{state(s)!maximally chaotic}\index{maximally chaotic state}
The maximally chaotic state $\chao{\Stset}$ of the convex set $\Stset$ is the baricenter of the set,
i.~e. it can be obtained by averaging over all pure states with the uniform measure, namely
\begin{equation}
\chao{\Stset}\doteq\int_{\Extr\Stset}\d\psi\,\psi
\end{equation}
where $\Extr\Stset$ denotes the set of extremal points of $\Stset$, and $\d\psi$ is the measure
that is invariant under isomorphisms of $\Stset$. 
\glossary{\Idx{convex2}$\Extr\Stset$ & extremal points of the convex set of states $\Stset$}
\end{definition}
From Definition \ref{def:chaostate} it follows that the maximally chaotic state is full-rank,
i. e. $\rnk[\chao{\Stset}]=\sqrt{\dim(\Stset)+1}$.  On the other hand, from Definition \ref{def:chaostate2} 
it follows that the group of isomorphisms of $\Stset$ leaves the state $\chao{\Stset}$ invariant
(but generally $\chao{\Stset}$ is not the only invariant state).
\section{Transformations and conditioned states}
\begin{grule}[Transformations form a monoid]\label{isemigroup}
\index{transformation!semigroup of}\index{transformation!composition}
\index{transformation!monoid of}
The composition $\tA\circ\tB$ of two transformations $\tA$ and $\tB$
is itself a transformation. Consistency of compostion of transformations requires {\em
associativity}, namely\index{transformation!associativity} 
\begin{equation}
\tC\circ(\tB\circ\tA)=(\tC\circ\tB)\circ\tA.
\end{equation}
There exists the identical transformation $\tI$ which leaves the physical system invariant, and
which for every transformation $\tA$ satisfies the composition rule 
\begin{equation}
\tI\circ\tA=\tA\circ\tI=\tA.
\end{equation}
Therefore, transformations make a semigroup with identity, i. e. a {\em monoid}.
\end{grule}
\begin{definition}[Independent systems and local experiments]\label{iindep}
  \index{independent systems}\index{independence}\index{experiment!local} We say that two physical
  systems are {\em independent} if on each system we can perform {\em local
experiments} that don't affect the other system for any joint state of the two systems. This can be
expressed synthetically with the commutativity of transformations of the local experiments, namely 
\begin{equation}
\tA^{(1)}\circ\tB^{(2)}=\tB^{(2)}\circ\tA^{(1)},
\end{equation}
where the label $n=1,2$ of the transformations denotes the system
undergoing the transformation.  
\end{definition}
In the following, when we have more than one independent system, we 
will denote local transformations as ordered strings of
transformations as follows
\begin{equation}
(\tA,\tB,\tC,\ldots)\doteq \tA^{(1)}\circ\tB^{(2)}\circ\tC^{(3)}\circ\ldots
\end{equation}
i. e. the transformation in parentheses corresponds to the local
transformation $\tA$ on system 1, $\tB$ on system 2, etc. 
\glossary{\Idx{transformations3a}$(\tA,\tB,\tC,\ldots)$ & local transformations}
\glossary{\Idx{transformations3b}$\tA^{(1)}\circ\tB^{(2)}\circ\tC^{(3)}\circ\ldots$ & local transformations}
\medskip
\begin{grule}[Bayes] When composing two transformations $\tA$ and $\tB$, the probability
$p(\tB|\tA)$ that $\tB$ occurs conditional on the previous occurrence of $\tA$ is given by the Bayes
rule\index{Bayes rule} 
\begin{equation}
p(\tB|\tA)=\frac{\omega(\tB\circ\tA)}{\omega(\tA)}.
\end{equation}
\end{grule}
The Bayes rule leads to the concept of {\em conditional
  state}:\index{state(s)!conditional}\index{conditional state}
\begin{definition}[Conditional state]\label{istatecond} The {\em conditional
state} $\omega_\tA$ gives the probability that a transformation
$\tB$ occurs on the physical system in the state $\omega$ after the
transformation $\tA$ has occurred, namely
\begin{equation}
\omega_\tA(\tB)\doteq\frac{\omega(\tB\circ\tA)}{\omega(\tA)}.
\end{equation}
\glossary{\Idx{state3}$\omega_\tA$ & conditional state (state $\omega$ conditioned by the
  transformation $\tA$)}
\end{definition}
\begin{remark}[Linearity of evolution]
  At this point it is worth noticing that the present definition of ``state'', which logically
  follows from the definition of experiment, leads to a {\em notion of evolution as
    state-conditioning}. In this way, each transformation acts linearly on the state space. In
  addition, since states are probability functionals on transformations, by dualism (equivalence
  classes of) transformations are linear functionals over the state space.  This clarifies the
  common misconception according to which it is impossible to mimic Quantum Mechanics as a mere classical
  probabilistic mechanics on a phase space viewed as a probability space since Quantum Mechanics
  admits linear evolutions only, whereas classical mechanics also admits nonlinear evolutions.
\end{remark}
\par In the following we will make extensive use of the functional notation
\begin{equation}
\omega_\tA\doteq\frac{\omega(\cdot\circ\tA)}{\omega(\tA)},
\end{equation}
where the centered dot stands for the argument of the map. Therefore, the notion of conditional state describes
the most general {\em evolution}. 
\bigskip
\par For the following it is convenient to extend the notion of state to that of {\em weight}, \index{weight} 
namely nonnegative bounded functionals $\tilde\omega$ over the set of transformations with
$0\le\tilde\omega(\tA)\le\tilde\omega(\tI)<+\infty$ for all transformations $\tA$.  To each weight
$\tilde\omega$ it corresponds the properly normalized state
\begin{equation}
\omega=\frac{\tilde\omega}{\tilde\omega(\tI)}.
\end{equation}
Weights make the convex cone $\tilde\Stset$ which is generated by the convex set of states $\Stset$.
\glossary{\Idx{convex}$\tilde\Stset$ & convex cone $\tilde\Stset$ generated by the convex set of
  states}
We are now in position to introduce the concept of operation.
\begin{definition}[Operation]\label{operation}\index{operation} 
To each transformation $\tA$ we can associate a linear map $\Op{\tA}:\;\Stset\longrightarrow\tilde\Stset$,
which sends a state $\omega$ into the unnormalized state $\tilde\omega_\tA\doteq
\Op{\tA}\omega\in\tilde\Stset$, defined
by the relation 
\begin{equation}
\tilde\omega_\tA(\tB)=\omega(\tB\circ\tA).
\end{equation}
\end{definition}
Similarly to a state, the linear form $\tilde\omega_\tA\in\tilde\Stset$ for fixed $\tA$ maps from
the set of transformations to the interval $[0,1]$. It is not strictly a state only due to lack of
normalization, since $0<\tilde\omega_\tA(\tI)\le 1$. The operation $\operatorname{Op}$ 
gives the conditioned state through the state-reduction rule \index{state!state-reduction} \index{state-reduction}
\begin{equation}
\omega_\tA=\frac{\tilde\omega_\tA}{\omega(\tA)}
\equiv\frac{\Op{\tA}\omega}{\Op{\tA}\omega(\tI)}.
\end{equation}
\glossary{\Idx{operation}$\Op{\tA}$ & operation corresponding to the transformation $\tA$}
\bigskip
\par The concept of conditional state naturally leads to the following category of transformations
\begin{definition}[Purity of transformations]\label{d:purtrans} A transformation is
called {\em pure} if it preserves purity of states, namely if
$\omega_\tA$ is pure for $\omega$ pure. \index{transformation!pure}\index{transformation!mixing}
\end{definition}
In contrast, we will call {\em mixing} a transformation which is not
pure. We will also call {\em pure} an action\index{action!pure} made only of pure
transformations and {\em mixing}\index{action!mixing} an action containing at least one mixing
transformation. 
\section{Dynamical and informational equivalence}\label{s:transequivalence}
From the Bayes rule, or, equivalently, from the definition of
conditional state, we see that we can have the following complementary situations:
\begin{enumerate}
\item there are different transformations which produce the same state
  change, but generally occur with different probabilities;
\item there are different transformations which always occur with the
  same probability, but generally affect a different state change.
\end{enumerate}
The above observation leads us to the following definitions of
dynamical and informational equivalences of transformations.
\begin{definition}[Dynamical equivalence of transformations]\label{d:dyneq}
\index{transformation!dynamical equivalence}
\index{dynamical equivalence of transformations}
 Two  transformations $\tA$ and $\tB$ are dynamically equivalent if 
$\omega_\tA=\omega_\tB$ for all possible states $\omega$ of the
system.  
\end{definition}
\medskip
\begin{definition}[Informational equivalence of transformations]
\index{transformation!informational equivalence}
\index{informational equivalence of transformations}
 Two transformations $\tA$ and $\tB$ are informationally equivalent if
$\omega(\tA)=\omega(\tB)$ for all possible states $\omega$ of the
system.  
\end{definition}
 \begin{definition}[Complete equivalence of transformations/experiments]
\index{transformation!complete equivalence}
\index{experiment!complete equivalence}
\index{complete equivalence!of transformations}
\index{complete equivalence!of experiments}
Two transformations/experiments are completely equivalent iff they are both
dynamically and informationally equivalent.
\end{definition}
Notice that even though two transformations are completely equivalent,
in principle they can still be different experimentally, in the sense
that they are achieved with different experimental apparatus. However, we emphasize that outcomes
in different experiments corresponding to equivalent transformations always provide
the same information on the state of  the object, and, moreover, the
corresponding transformations of the state are the same.
\section{Informational compatibility}
\index{transformation!coexistence}
\index{transformation!informational compatibility}
\index{informational compatibility of transformations}
\index{coexistence of transformations}
The concept of dynamical equivalence of transformations leads one to
introduce a convex structure also for transformations. We first need
the notion of informational compatibility.
\begin{definition}[Informational compatibility or coexistence] We say that
two transformations $\tA$ and $\tB$ are {\em coexistent} or {\em
informationally compatible} if one has 
\begin{equation}
\omega(\tA)+\omega(\tB)\le 1,\quad\forall\omega\in\Stset,\label{compatible}
\end{equation}
\end{definition}
The fact that two transformations are coexistent means that, in principle, they can occur in the
same experiment, namely there  
exists at least an action containing both of them. We have named the
present kind of compatibility "informational" since it is actually
defined on the informational equivalence classes of transformations.
Notice that the relation of coexistence is symmetric, but is not
reflexive, since a transformation can be coexistent with itself only 
if $\omega(\tA)\le 1/2$. The present notion of coexistence is the
analogous of that introduced by Ludwig \cite{Ludwig-axI} for the
"effects". This notion is also related to that of "exclusive"
transformations, since they correspond to exclusive outcomes [see also
Ref. \cite{Kraus74} in regards "exclusive" implies "coexistent", but
generally not the reverse].
\par We are now in position to define the "addition" of coexistent transformations.
\begin{grule}[Addition of coexistent transformations]\label{g:addtrans}
\index{transformation!addition} For any two
  coexistent transformations $\tA$ and $\tB$  we define the
  transformation $\tS=\tA_1+\tA_2$ as the transformation corresponding
  to the event $e=\{1,2\}$, namely the apparatus signals that either
  $\tA_1$ or $\tA_2$ occurred, but doesn't specify which one.
By definition, one has the distributivity rule 
\begin{equation}\label{r:sum1}
\forall\omega\in\Stset\qquad\omega(\tA_1+\tA_2)=\omega(\tA_1)+\omega(\tA_2),
\end{equation}
whereas the state conditioning is given by
\begin{equation}\label{r:sum2}
\forall\omega\in\Stset\qquad
\omega_{\tA_1+\tA_2}=\frac{\omega(\tA_1)}{\omega(\tA_1+\tA_2)}
\omega_{\tA_1}+\frac{\omega(\tA_2)}{{\omega(\tA_1+\tA_2)}}\omega_{\tA_2}.
\end{equation}
\glossary{\Idx{transformations5}$\tA+\tB$ & addition of compatible transformations}
\end{grule}
Notice that the two rules in Eqs.  (\ref{r:sum1}) and (\ref{r:sum2}) completely specify the 
transformation $\tA_1+\tA_2$, both informationally and dynamically. Eq. (\ref{r:sum2}) can be more
easily restated in terms of operations as follows:
\begin{equation}
\forall\omega\in\Stset\qquad
\Op{\tA_1+\tA_2}\omega=\Op{\tA_1}\omega+\Op{\tA_2}\omega.
\end{equation}
Addition of compatible transformations is the core of the description
of partial knowledge on the experimental apparatus. Notice also that
the same notion of coexistence can extended to "propensities" as well
(see Definition \ref{d:propensity}). 
\begin{definition}[Indecomposable transformation] We call a transformation $\tT$ {\em
    indecomposable}, if there are no coexistent transformations summing to it.
\end{definition}
From the above definition we can see that the equivalent of quantum unitary transformations could be
defined in terms of indecomposable isometric transformations. 
\begin{grule}[Multiplication of a transformation by a scalar]\label{g:scalmult}
\index{transformation!multiplication by a scalar}
For each transformation $\tA$ the transformation $\lambda\tA$ for
$0\le\lambda\le 1$ is defined as the transformation which is 
dynamically equivalent to $\tA$, but which occurs with probability
$\omega(\lambda\tA)=\lambda\omega(\tA)$.   
\end{grule}
\begin{remark} [No-information from identity transformations]\label{r:noinfo}
\index{transformation!identity}\index{no-information from id. transformation}
At this point a warning is in order, as regards the transformations that
are dynamically equivalent to the identity, namely the {\em
probabilistic identity transformations}. According to the Rule
\ref{g:scalmult} for multiplication of transformations by a scalar, 
a probabilistic identity transformation will be of the form $p\tI$,
where $p$ is the probability that the transformation occurs, namely
$p=\omega(p\tI)$. One could now imagine an hypothetical situation of
a "classical" experiment which leaves the object identically undisturbed,
independently of its state, but still with many different outcomes $j$
that are signaled by the apparatus. If such an experiment had
an action of the form $\AA=\{p_j\tI\}$, it would provide no
information on the state $\omega$ of the object, since by definition the
probabilities of the outcomes will be independent on $\omega$, because
$\omega(p_j\tI)=p_j$. Therefore, a "classical" experiment makes sense
only for an action $\AA=\{\tA_j\}$ made of non-identical transformations,
but with the set of states restricted to be all invariant under $\AA$.
\end{remark}
\bigskip
It is now natural to introduce a norm over transformations as follows.
\begin{theorem}[Norm for transformations]\label{t:transnorm} The following quantity
\index{transformation!norm}
\glossary{\Idx{transformations4}$\n{\tA}$ & norm of transformation}
\begin{equation}
\n{\tA}=\sup_{\omega\in\Stset}\omega(\tA),\label{norm}
\end{equation}
is a norm on the set of transformations. In terms of such norm
all transformations are contractions.
\index{transformation!contraction}
\end{theorem}
\Proof The quantity in Eq. (\ref{norm}) satisfy the sub-additivity relation
$\n{\tA+\tB}\le\n{\tA}+\n{\tB}$, since
\begin{equation}
\n{\tA+\tB}=\sup_{\omega\in\Stset}[\omega(\tA)+\omega(\tB)]\le
\sup_{\omega\in\Stset}\omega(\tA)+\sup_{\omega'\in\Stset}\omega'(\tB)=
\n{\tA}+\n{\tB}.
\end{equation}
Moreover, it obviously satisfies the identity
\begin{equation}
\n{\lambda\tA}=\lambda\n{\tA}.
\end{equation}
It is also clear that, by definition, for each transformation $\tA$ one
has $\n{\tA}\le1$, namely transformations are contractions.\qed
Obviously the multiplication of a transformation $\tA$ by a
scalar is more generally defined for a scalar $\lambda\le\n{\tA}^{-1}$,
which can be larger than unity.  In terms of the norm (\ref{norm}) one can
equivalently define coexistence (informational compatibility) using
the following corollary
\begin{corollary} Two transformations $\tA$ and $\tB$ are
  coexistent iff $\tA+\tB$ is a contraction.
\end{corollary}
\Proof If the two transformations are coexistent, then
from Eqs. (\ref{compatible}) and (\ref{norm}) one has that
$\n{\tA+\tB}\le 1$. On the other hand, if $\n{\tA+\tB}\le 1$, this
means that Eq. (\ref{norm}) is satisfied for all states, namely the
transformations are coexistent.\qed 
\begin{corollary} The transformations $\lambda\tA$ and
  $(1-\lambda)\tB$ are compatible for any couple of transformations
  $\tA$ and $\tB$. 
\end{corollary}
\Proof Clearly
$\n{\lambda\tA+(1-\lambda)\tB}\le\lambda\n{\tA}+(1-\lambda)\n{\tB}\le 1$.\qed 
\medskip
\par The last corollary implies the rule
\begin{grule}[Convex structure of transformations]\label{r:convextrans}
\index{transformation!convex structure}
Transformations form a convex set, namely for any two 
transformations $\tA_1$ and $\tA_2$ we can consider the transformation
$\tA$ which is the {\em mixture} of $\tA_1$ and $\tA_2$ with
probabilities $\lambda$ and $1-\lambda$. Formally, we write
\begin{equation}
\tA=\lambda\tA_1+(1-\lambda)\tA_2,\quad 0\le\lambda\le 1,\label{ala}
\end{equation}
with the following meaning: the transformation $\tA$ is itself a
probabilistic transformation, occurring with overall probability
\begin{equation}
\omega(\tA)=\lambda\omega(\tA_1)+(1-\lambda)\omega(\tA_2),
\end{equation}
meaning that when the transformation $\tA$ occurred we
know that the transformation dynamically was either $\tA_1$ with 
(conditioned) probability $\lambda$ or $\tA_2$ with
probability $(1-\lambda)$.
\end{grule}
We have seen that the transformations form a convex set, more 
specifically, a spherically truncated convex cone, namely we can
always add transformations or multiply a transformation by a positive scalar if the result is a contraction. In
the following we will denote the spherically truncated convex cone of transformations 
as $\Trnset$. 
\glossary{\Idx{convex3}$\Trnset$ & truncated convex cone of transformations}
\medskip\par We should be aware that extremality of transformations in
relation to their convex structure is not equivalent to the concept of
purity in Definition \ref{d:purtrans}, since a pure transformation is not necessarily 
extremal (just consider the convex combination of two different transformations that map to the same
pure state), and vice-versa the fact that a transformation is mixing doesn't logically imply that it can
be always regarded as a convex combination of extremal transformations.  
\medskip
\begin{remark}[Banach algebra of transformations] The convex cone of transformations can be extended
(on the embedding affine space) to a real Banach algebra equipped with the norm given in Theorem
\ref{t:transnorm}, the closure corresponding to an approximation criterion for transformations.
\end{remark}
\medskip
\par An obvious consequence of the rule \ref{r:convextrans} is that actions too form a convex set, namely
\begin{grule}[Convex structure of actions]\index{action!convex structure}
 Actions make a convex set,
namely for any two actions $\AA=\{\tA_j\}$ and $\AB=\{\tB_j\}$ we can
consider the action $\AC$ which is the {\em mixture} of $\AA$ and $\AB$ with
probabilities $\lambda$ and $1-\lambda$
\begin{equation}
\AC=\lambda\AA+(1-\lambda)\AB=\{\lambda\tA_j,(1-\lambda)\tB_i\},
\quad 0\le\lambda\le 1,\label{AlA},
\end{equation}
with the following meaning: the action $\AC$ has the union of outcomes
of actions $\AA$ and $\AB$, and contains the transformations
$\lambda\tA_j$ and $(1-\lambda)\tB_j$ which are dynamically equivalent
to those of actions $\AA$ and $\AB$.
\glossary{\Idx{convex4}$\Actset$ & convex set of actions}
\end{grule}
\section{Propensities}
Informational equivalence allows one to define equivalence classes of transformations, which we may
want to call {\em propensities}, since they give the occurrence probability of a transformation for
each state, \ie its ``disposition'' to occur.  
\begin{definition}[Propensities]\label{d:propensity}
\index{propensity} We call {\bf propensity} an
  informational equivalence class of transformations.  
\end{definition}
It is easy to see that the present notion of propensity corresponds
closely to the notion of "effect" introduced by Ludwig
\cite{Ludwig-axI}. However, we prefer to keep a separate word, since
the "effect" has been identified with a quantum mechanical notion and
a precise mathematical object (\ie a positive contraction).
\glossary{\Idx{propensity1}$\cA,\cB,\ldots,$ & propensities}
\glossary{\Idx{propensity2}$[\tA]$ & propensity containing the transformation $\tA$}
In the following we will denote propensities with underlined symbols
as $\cA$, $\cB$, etc., and we will use the notation $[\tA]$ for the
propensity containing the transformation $\tA$, and also write
$\tA'\in[\tA]$ to say that $\tA'$ is informationally equivalent to
$[\tA]$. It is clear that $\lambda\tA$ and $\lambda\tB$ belong to the
same equivalence class iff $\tA$ and $\tB$ are informationally
equivalent. This means that also for propensities multiplication
by a scalar can be defined as $\lambda[\tA]=[\lambda\tA]$. Moreover,
since for $\tA'\in[\tA]$ and $\tB'\in[\tB]$ one has 
$\tA'+\tB'\in[\tA+\tB]$, we can define addition of propensities as
\index{propensity!addition}
$[\tA]+[\tB]=[\tA+\tB]$ for any choice of representatives $\tA$ and
$\tB$ of the two added propensities. Also, since all transformations
of the same equivalence class have the same norm, we can extend
the definition (\ref{norm}) to propensities as $\n{[\tA]}=\n{\tA}$ for
any representative $\tA$ of the class. It is easy to check
sub-additivity on classes, which implies that it is indeed a norm. In
fact, one has 
\begin{equation}
\n{[\tA]+[\tB]}=\n{\tA+\tB}\le\n{\tA}+\n{\tB}=\n{[\tA]}+\n{[\tB]}.
\end{equation}
Therefore, it follows that also propensities form a spherically
truncated convex cone (which is a convex set), and in the following we
will denote it by $\Cntset$.
\glossary{\Idx{convex5}$\Cntset$ & truncated convex cone of propensities}
\medskip
\par With the present norm for propensities, Ludwig \cite{Ludwig-axI}
introduces the notion of "ensembles with maximal absorption",
corresponding to the state achieving the norm of the propensity
$l(\omega)=\n{l}$ and of "ensembles totally absorbed" when
$l(\omega)=1$.  
\glossary{\Idx{propensity3}$l$ & propensity}
\begin{remark}[Duality between the convex sets of states and of propensities]
From the Definition \ref{istate} of state it follows that the
convex set of states $\Stset$ and the convex sets of propensities
$\Cntset$ are dual each other, and the latter can be regarded as the
set of positive linear contractions over the set of states, namely the
set of positive functionals $l$ on $\Stset$ with unit upper bound, and
with the functional $l_{[\tA]}$ corresponding to the propensity $[\tA]$
being defined as
\glossary{\Idx{propensity4}$l_{[\tA]}$ & propensity containing the transformation $\tA$}
\begin{equation}
l_{[\tA]}(\omega)\doteq\omega(\tA).
\end{equation}
In the following we will often identify propensities with their
corresponding functionals, and denote them by lowercase letters
$a,b,c,\ldots$, or $l_1,l_2,\ldots$. Finally, notice that the notion of
coexistence (informational compatibility) extends naturally to
propensities. 
\end{remark}
\begin{remark}[Dual cone notation]
We can write the propensity linear functionals on $\Stset$ with the equivalent pairing notations 
\begin{equation}
l_{\cA}(\omega)\doteq\omega(\cA)\equiv(\cA,\omega).
\end{equation}
\end{remark}
\begin{definition}[Observable]\index{observable} We call 
  observable a set of propensities $\AL=\{l_i\}$ which is informationally equivalent to an action
  $\AL\in\underline{\AA}$, namely such that there exists an action $\AA=\{\tA_j\}$ for of which one
  has $l_i\in\underline{\tA_j}$.
\end{definition}
Clearly, the generalized observable is normalized to the constant unit functional, i. e. $\sum_il_i=1$.
\begin{definition}[Informationally complete observable] An observable
  $\AL=\{l_i\}$ is informationally complete if each propensity can be written as a linear combination of
  the of elements of $\AL$, namely for each propensity $l$ there exist coefficients $c_i(l)$ such that
\begin{equation}
l=\sum_ic_i(l)l_i.
\end{equation}
\end{definition}
Clearly, using an informationally complete observable one can reconstruct any state
$\omega$ from just the probabilities $l_i(\omega)$, since one has
\begin{equation}
\omega(\tA)=\sum_ic_i(l_{\underline{\tA}})l_i(\omega).
\end{equation}
\begin{grule}[Partial ordering between propensities] For two
propensities $l_1, l_2\in\Cntset$ we write  $l_1\le l_2$ when
$l_1(\omega)\le l_2(\omega)$ $\forall\omega\in\Stset$.
\end{grule}
In Ref. \cite{Ludwig-axI} the present partial ordering is interpreted
saying that $l_2$ is {\em more sensitive} than $l_1$.
\section{Dynamic compatibility}
\par Regarding the dynamical face of the concept of "transformation", we
can introduce another notion of compatibility, which is closer to the one
usually considered in quantum mechanics. 
\begin{definition}[Dynamical compatibility]
\index{transformation!dynamic compatibility}
\index{dynamic compatibility!of transformations}
 We say that two transformations $\tA$ and $\tB$ are dynamically compatible if they commute, namely
$\tA\circ\tB=\tB\circ\tA$. 
\end{definition}
An example of dynamically compatible transformations is provided by a
couple of local transformations on independent object systems. 
\section{Compatibility of experiments}
The concept of dynamical compatibility naturally extends to actions as
follows.
\begin{definition}[Compatible experiments]
\index{experiment!dynamic compatibility}
\index{dynamic compatibility!of experiments}
 We call two experiments made with two different apparatuses compatible---i. e. they can be
  performed contextually on the same object system---when their
  relative order is irrelevant, namely their action are made of
  transformations that are dynamically compatible.
\end{definition}
The above definitions means that the actions $\AA=\{\tA_j\}$ and
$\AB=\{\tB_i\}$ of two compatible experiments are such that
$\tA_j\circ\tB_i=\tB_i\circ\tA_j$ for all transformations of $\AA$ and
$\AB$. This allows one to define the contextually joint experiment, with action
$\AC=\AA\&\AB$ and $\AC=\{\tC_{ij}\}$, where now the possible outcomes
are the product events $ij$ corresponding to transformations
$\tC_{ij}=\tA_j\circ\tB_i\equiv\tB_i\circ\tA_j$. Notice that when joining
contextually two experiments, generally their outcomes are correlated,
namely $\omega(\tB_i\circ\tA_j)\neq\omega(\tB_i)\omega(\tA_j)$, and
compatibility only implies the identity
\begin{equation}
\frac{\omega_{\tA_j}(\tB_i)}{\omega_{\tB_i}(\tA_j)}=\frac{\omega(\tB_i)}{\omega(\tA_j)}.
\end{equation}
The present definition of contextuality may look artificial, but it is in line with the
"a-temporal" scenario of our definition of experiment, where "time" refers only to the before-after
ordering between the action---the "cause"---and the transformation of the object system---the
"effect". In this fashion, the only logical way of defining contextually joint experiments is to
consider them as equivalent for any choice of their ordering. Clearly, in any practical
definition of contextual joint experiments, at least we need to have the apparatuses as independent
systems themselves. On the other hand, for incompatible experiments with actions $\AA$ and $\AB$
one can always define the experiment corresponding to the cascade of the previous two on the same
object system, with action $\AB\circ\AA=\{\tB_i\circ\tA_j\}$.

Notice how the present definition of compatible experiments is deeply related to that of independent
systems. Indeed, if there exists a nonempty commutant for a complete set of transformations, this
will allow one to define two subsystems, at least in the sense of ``virtual subsystems'' \cite{zanardi01}.

The informational counterpart of compatible experiments will be the following

\begin{definition}[Informational compatibility of experiments]
\index{experiment!informational compatibility}
\index{informational compatibility!of experiments}
We say that two experiments with actions $\AA=\{\tA_j\}$ and
$\AB=\{\tB_i\}$ are informationally compatible when there exists a
third experiment whose action $\AC$ has marginals informationally
equivalent to $\AA$ and $\AB$, namely we can partition the outcomes in
such a way that we can write $\AC=\{\tC_{ij}\}$ with
$\sum_i\tC_{ij}\in[\tA_j]$  and $\sum_j\tC_{ij}\in[\tB_i]$. 
\end{definition}
Notice that dynamically compatible experiments are always
informationally compatible, since one has 
\begin{equation}
\begin{split}
\sum_i\omega(\tB_i\circ\tA_j)=&\sum_i\omega(\tA_j)\omega_{\tA_j}(\tB_i)\equiv\omega(\tA_j),\\
\sum_j\omega(\tB_i\circ\tA_j)=&\sum_j\omega(\tA_j\circ\tB_i)=\sum_j\omega(\tB_i)\omega_{\tB_i}(\tA_j)\equiv\omega(\tB_i),
\end{split}
\end{equation}
whereas, generally, for the cascade of experiments
$\AB\circ\AA=\{\tB_i\circ\tA_j\}$, one has only 
$\sum_i\tB_i\circ\tA_j\in[\tA_j]$ , but generally $\sum_j\tB_i\circ\tA_j\not\in[\tB_i]$.  
\section{Predictability and distances between states}
\begin{definition}[Predictability and resolution]\label{def:res} 
\index{transformation!predictable}\index{predictable!transformation}
\index{propensity!predictable}\index{predictable!propensity}
We will call a transformation $\tA$---and likewise its
propensity---{\em predictable} if there exists a state for which
$\tA$ occurs with certainty and some other state for which it never
occurs. The transformation (propensity) will be also called {\em
resolved} if the state for which it occurs with certainty is
unique---whence pure.
An action will be called {\em predictable} when it is made only
of predictable transformations, and {\em resolved} when all
transformations are resolved.
\end{definition}
The present notion of predictability for propensity corresponds to that of
"decision effects" of Ludwig \cite{Ludwig-axI}. For a predictable
transformation $\tA$ one has $\n{\tA}=1$. Notice that 
a predictable transformation is not deterministic, and it can
generally occur with nonunit probability on some state $\omega$. 
Predictable propensities $\tA$ correspond to affine functions
$f_\tA$ on the state space $\Stset$ with $0\le f_\tA\le 1$ achieving
both bounds. Their set will be denoted by $\Prdset$. 
\par Via propensities, we can also introduce 
notions of {\em distance} and of  {\em orthogonality} on the state space $\Stset$.
\glossary{\Idx{convex6}$\Prdset$ & convex set of predictable propensities}
\begin{definition}[Distance between states]\label{d:state-dist} 
\index{state(s)!metric}\index{metric!for states}\index{distance between states}
Let $\Cntset$ denote the set of propensities on the convex set of states $\Stset$. Define the
"distance" between states $\omega,\zeta\in\Stset$ as follows
\begin{equation}
d(\omega,\zeta)=\sup_{l\in\Cntset}l(\omega)-l(\zeta).\label{state-dist}
\end{equation}
\end{definition}
\begin{theorem} The function (\ref{state-dist}) is a metric on $\Stset$.
\end{theorem}
\Proof For every propensity $l$, $1-l$ is also a propensity, whence
\begin{equation}
d(\omega,\zeta)=\sup_{l\in\Cntset}(l(\omega)-l(\zeta))=
\sup_{l'\in\Cntset}((1-l')(\omega)-(1-l')(\zeta))=
\sup_{l'\in\Cntset}(l'(\zeta)-l'(\omega))=d(\zeta,\omega),
\end{equation}
namely $d$ is symmetric. On the other hand, $d(\omega,\zeta)=0$ implies that $\zeta=\omega$, since
the two states must give the same probabilities for all transformations. Finally, one has
\begin{equation}
d(\omega,\zeta)=\sup_{l\in\Cntset}(l(\omega)-l(\theta)+l(\theta)-l(\zeta))\le
\sup_{l\in\Cntset}(l(\omega)-l(\theta))+
\sup_{l\in\Cntset}(l(\theta)-l(\zeta))=d(\omega,\theta)+d(\theta,\zeta),
\end{equation}
namely it satisfy the triangular inequality, whence $d$ is a metric.\qed
\par One can see that, by construction, the distance is bounded as
$d(\omega,\zeta)\le 1$, since the maximum value of $d(\omega,\zeta)$
is achieved for $l(\omega)=1$ 
and $l(\zeta)=0$. Moreover, since for a linear function on a convex
domain both maximum and minimum are achieved on facets (\ie convex
hulls of some extremal points), this means that the bound
$d(\omega,\zeta)=1$ can be achieved only when $\omega$ and $\zeta$  
lie on different facets of the convex set. Finally, for convex combinations we have
the following
\begin{lemma}
Mixing reduces distances linearly.
\end{lemma}
\Proof
For any convex combination $\theta=\alpha\omega+(1-\alpha)\zeta$ one has
$d(\theta,\zeta)=\alpha d(\omega,\zeta)$, since
\begin{equation}
\begin{split}
d(\theta,\zeta)=\sup_{l\in\Cntset}(\alpha l(\omega)+(1-\alpha)l(\zeta)-l(\zeta))=
\sup_{l\in\Cntset}(\alpha l(\omega)-\alpha l(\zeta))=\alpha d(\omega,\zeta).
\end{split}
\end{equation}
\begin{definition}[Orthogonality of states]\label{d:orthostates} Two states
$\omega,\zeta\in\Stset$ are called {\em orthogonal} (denoted as  
$\omega\perp\zeta$) if $d(\omega,\zeta)=1$. 
\end{definition}
\begin{definition}[Metrical dimensionality] The metric dimensionality is the maximum number of pairwise
orthogonal states according to Definition \ref{d:orthostates}.\index{dimension!metrical}
\glossary{\Idx{dimension4}$\mdim{\Stset}$ & metrical dimension of the convex set of states $\Stset$}
\end{definition}
For example, the metric dimensionality of any $N$-hypersphere is $2$, since the set of predictable
propensity is made of the linear functions $f_{\vec m}(\vec n)=\tfrac{1}{2}(1+\vec n\cdot\vec m)$
where $\vec m$ is a unit vector, and the metric is $d(\vec n,\vec
n')=\max_{\vec m}\tfrac{1}{2}\vec m\cdot (\vec n-\vec n')\equiv
\tfrac{1}{2}|\vec n-\vec n'|$, whence one sees that  only antipodal points have distance $1$.
\begin{example}
Consider the trace-norm distance on the convex set of density
operators over the Hilbert space $\sH$ $d(x,y)=\frac{1}{2}\n{x-y}_1$. For pure states
one has $d(x,y)=\sqrt{1-|\<\psi_x|\psi_y\>|^2}$. Therefore, the metric
structure of $\sH$ is rediscovered via the inner metric of the
state-space, and orthogonality in $\sH$ means maximal inner
distance $d(x,y)=1$ in the state space.
\end{example}
\begin{definition}[Isometric transformations] A transformation $\tU$ is called {\em isometric} if it
  preserves the distance between states, namely 
\begin{equation}
d(\omega_\tU,\zeta_\tU)\equiv d(\omega,\zeta),\qquad \forall \omega,\zeta\in\Stset.
\end{equation}
\end{definition}
Isometric transformations are isomorphisms of the convex of states $\Stset$.
On the other hand, isomorphisms of the convex set of propensities $\Cntset$ are also isometric
transformations of states, since
\begin{equation}
\sup_{l_\tA\in\Cntset}\omega(\tA\circ\tU)-\zeta(\tA\circ\tU)
=\sup_{l_{\tA\circ\tU}\in\Cntset}\omega(\tA)-\zeta(\tA)=d(\omega,\zeta).
\end{equation}

\begin{definition}[Perfectly discriminable set of states] We call a set of states $\{\omega_n\}_{n=1,N}$
  {\em perfectly discriminable} if there exists an action $\AA=\{\tA_j\}_{j=1,N}$ with transformations
  $\tA_j\in l_j$ corresponding to a set of  predictable propensities $\{l_n\}_{n=1,N}$ satisfying the 
  relation \index{state(s)!perfectly discriminable}
\begin{equation}
l_n(\omega_m)=\delta_{nm}.
\end{equation}
\end{definition}
\begin{definition}[Informational dimensionality] We call the {\em informational dimension}
\index{dimension!informational} of the convex set of states $\Stset$, denoted by $\idim{\Stset}$,
the maximal cardinality of perfectly discriminable set of states in $\Stset$.
\glossary{\Idx{dimension3}$\idim{\Stset}$ & informational dimension of the convex set of states $\Stset$}
\end{definition}
\begin{theorem}
Two orthogonal states are perfectly discriminable.
\end{theorem}
\Proof If the two states, say $\omega_1$ and $\omega_2$, are orthogonal, then this means that
$1=d(\omega_1,\omega_2)=\sup_{l\in\Cntset}(l(\omega_1)-l(\omega_2))$, namely there  exists a 
propensity $l_1$ such that $l_1(\omega_1)=1$ and $l_1(\omega_2)=0$. Now, consider the propensity
$l_2=1-l_1$,  and this will satisfy by definition $l_2(\omega_1)=0$ and $l_2(\omega_2)=1$. Now,
construct an apparatus with action $\AA=\{\tA_1,\tA_2\}$, with $\tA_n\in l_n$, for $n=1,2$, and you
are done. 
\begin{remark}\label{r:discriminable} Note that it seems that the above theorem doesn't generalize
  to more than two mutually orthogonal states. In fact, if there are $N>2$ states that are
  orthogonal to each other, then we only know that for each of the $\frac{1}{2}N(N-1)$ couples of
  states, say $\zeta_1$ and $\zeta_2$, there exists a predictable propensity $l$ for which
  $l(\zeta_1)=1$ and $l(\zeta_2)=0$.  This does not even guarantee that if a state $\omega$ is
  orthogonal to both $\zeta_1$ and $\zeta_2$, then it should be orthogonal also to any their convex
  linear combination. In fact, orthogonality implies the existence of two propensities $l_1$ and $l_2$
  such that $l_1(\omega)=l_2(\omega)=1$ and $l_1(\zeta_1)=l_2(\zeta_2)=0$. Now, the distance of
  $\omega$ from the convex combination $\alpha\zeta_2+(1-\alpha)\zeta_1$ is given by
\begin{equation}
d(\omega,\alpha\zeta_2+(1-\alpha)\zeta_1)=\sup_{l\in\Prdset}
[l(\omega)-\alpha l(\zeta_2)-(1-\alpha)l(\zeta_1)]=\sup_{l\in\Prdset}
\alpha[l(\omega)-l(\zeta_2)]+(1-\alpha)[l(\omega)-l(\zeta_1)],
\end{equation}
which is equal to one if and only if one has both $l(\zeta_2)=l(\zeta_1)=0$. Therefore, in order to
preserve orthogonality for convex combination, we need a functional achieving $l(\omega)=1$, and for
which $l(\zeta)=0$ for all states $\zeta\perp\omega$: it seems that the existence of such
functional is not implied by the existence of many functionals $l_\zeta$, with $l_\zeta(\zeta)=1$
and $l_\zeta(\omega)=0$ for all states $\omega\perp\zeta$. Also convex combination of the propensities
doesn't help. In fact, consider a linear combination of the propensities $h=\beta l
_{\zeta_1}+(1-\beta)l_{\zeta_2}$ on the mixture $\alpha\zeta_1+(1-\alpha)\zeta_2$. One has
$h[\alpha\zeta_1+(1-\alpha)\zeta_2]=\beta(1-\alpha)l_{\zeta_1}(\zeta_2)+(1-\beta)\alpha
l_{\zeta_2}(\zeta_1)$ which we want to vanish for all $\alpha$, giving the following value for $\beta$
\begin{equation}
\beta=\frac{\alpha l_{\zeta_2}(\zeta_1)}{\alpha l_{\zeta_2}(\zeta_1)-(1-\alpha)l_{\zeta_1}(\zeta_2)},
\end{equation}
which not necessarily satisfies $0\le\beta\le 1$. 
\end{remark}
\par The above considerations lead us to restrict the notion of joint orthogonality \index{joint
  orthogonality} as follows
\begin{definition}[Joint orthogonality]\label{def:jort} We say that a set of states $\set{S}$  is
  {\em jointly orthogonal}
  to a given state $\omega$ if each state of their convex hull $\Conv(\set{S})$  is orthogonal to
  $\omega$. 
\end{definition}
\glossary{\Idx{convexhull}$\Conv(\set{S})$ & convex hull of the set $\set{S}$}
\glossary{\Idx{convexhull}$\Stset_\omega^\perp$ & convex set of states jointly orthogonal to $\omega$}
\glossary{\Idx{convexhull}$\Stset_\set{S}^\perp$ & convex set of states jointly orthogonal to the
  set of states $\set{S}$}
Clearly, the definition of joint orthogonality to a state extends to joint orthogonality to a
(convex) set of states. We will denote the convex set of states in $\Stset$ jointly orthogonal to $\omega$ by
  $\Stset_\omega^\perp$, and the convex set of states in $\Stset$ jointly orthogonal to the set $\set{S}$ by
  $\Stset_\set{S}^\perp$.
\par Definition \ref{def:jort} is also equivalent to
\begin{theorem}
A state $\omega$ is jointly orthogonal to a set of states $\set{S}$ if and only if there exists a
predictable propensity $l$ achieving $l(\omega)=1$ and which vanishes identically over the whole set $\set{S}$.
\end{theorem}
The above theorem also implies the following corollary
\begin{corollary} Any set $\Stset_\set{S}^\perp$ is a planar section of $\Stset$. 
\end{corollary}
\begin{definition}[Discriminating observable]
  \index{observable!discriminating}\index{propensity!observable} An observable
  $\AL=\{l_j\}$ is {\em discriminating for} $\Stset$ when it discriminates a set of
  states with cardinality equal to the informational dimension $\idim{\Stset}$ of $\Stset$.
\end{definition}
\bigskip
\begin{remark} It is natural to conjecture that a resolved predictable action (see Definition \ref{def:res})
  is the same as a discriminating observable. In fact, by definition, each transformation of a
  resolved predictable action must be predictable. On the 
  other hand, if it is not resolved, then there will be at least an unresolved transformation, which will occur with
probability one for at least two different states. These states could in principle be resolved by  another 
transformation, but there is no guarantee that such transformation exists. Therefore, it is not 
obvious whether the cardinality of all resolved predictable actions are the same, whence it would
coincide with $\idim{\Stset}$. 
\end{remark}
\begin{remark}[Different dimensionalities for $\Stset$]\index{dimension!relations between different types}
We have introduced three different dimensionalities for the convex set of states $\Stset$: 1) the
Caratheodory's dimension $\cdim{\Stset}$; 2) the metrical dimension $\mdim{\Stset}$; and 3) the
informational  dimension $\idim{\Stset}$. In Quantum Mechanics they all coincide. However, in
general it seems that there are no definite reasons why they should have the same value. Let's
analyze the possible relation between different definitions. 
\par In order to establish a relation between Caratheodory's  and metrical dimensionalities, one
should first establish if: (a) for any state there always exists a minimal convex decompositions into
pure states that are pairwise orthogonal; (b) any convex combination of pairwise orthogonal states
is minimal for the resulting mixed state. Clearly, assertion (a) would imply that the maximal
rank of a state is smaller than the maximal number of pairwise orthogonal states, namely:
$\cdim{\Stset}\le\mdim{\Stset}$.  On the other hand, assertion (b) would imply that $\mdim{\Stset}$ is
the maximal rank of a state, whence the two dimensions coincide, i.~e. $\mdim{\Stset}=\cdim{\Stset}$. 
\par As regards a relation between informational and metrical dimensionalities, we have noticed in
Remark \ref{r:discriminable} that pairwise orthogonal states are not necessarily discriminable,
whereas, obviously the reverse is true, namely perfectly discriminable states are pairwise
orthogonal. Therefore, the maximal number of perfectly discriminable states is bounded by the
maximal number of pairwise orthogonal states, whence $\idim{\Stset}\le\mdim{\Stset}$.
\end{remark}
\section{Local state}\index{local!state}
\begin{definition}[Local state]\label{istateloc}
\index{state(s)!local}\index{local!state}
In the presence of many independent systems in a joint state $\Omega$, we define the {\bf
local state} $\omega^{(n)}$ of the $n$-th system the state that gives the probability for any local
transformation $\tA$ on the $n$-th system, with all other systems
untouched, namely
\begin{equation}
\omega^{(n)}(\tA)\doteq\Omega(\tI,\ldots,\tI,\underbrace{\tA}_{n\text{th}},\tI,\ldots).
\end{equation}
\glossary{\Idx{state4}$\omega^{(n)}$ & local state}
\end{definition}
For example, for two systems only, (which is equivalent to group
$n-1$ systems into a single one), we just write $\omega^{(1)}(\tA)=\Omega(\tA,\tI)$.
Notice that generally the commutativity Rule \ref{iindep} doesn't
imply that the occurrence of a transformation $\tB$ on system 2
doesn't change the probability of occurrence of any other
transformation $\tA$ on system 1, namely, generally 
\begin{equation}
\tA^{(1)}\circ\tB^{(2)}=\tB^{(2)}\circ\tA^{(1)}\not\Longrightarrow
\frac{\Omega(\cdot,\tB)}{\Omega(\tI,\tB)}=\Omega(\cdot,\tI).
\end{equation}
In other words, the occurrence of the transformation $\tB$ on system 2
generally affects the conditioned local state on system 1, namely
one has
\begin{equation}
\Omega_{\tB^{(2)}}(\cdot,\tI)\doteq\frac{\Omega(\cdot,\tB)}{\Omega(\tI,\tB)}\neq
\Omega(\cdot,\tI)\equiv\omega^{(1)}.\label{nonlocalp}
\end{equation}
Therefore, in order not to violate the relativity principle, for
independent systems (\eg space-like separated) we need to require
explicitly the acausality principle:
\begin{grule}[Acausality of local transformations]\label{iacausal}
\index{transformation!local}\index{local!transformation}
\index{acausality of transformation}
Any local action on a system is equivalent to the identity
transformation when viewed from an independent system, namely, 
in terms of states one has
\begin{equation}
\forall\AA\qquad
\sum_{\tA_j\in\AA}\Omega(\cdot,\tA_j)=\Omega(\cdot,\tI)\equiv\omega^{(1)}
\end{equation}
\end{grule}
The acausality of local transformations Rule \ref{iacausal} along
with the existence of inequivalent actions imply the existence of
indistinguishable incompatible mixtures. 
\begin{corollary}[Existence of equivalent incompatible mixtures]\label{iineqmix} 
\index{equivalent incompatible mixtures}\index{mixtures!incompatible}\index{mixtures!equivalence of}
For any two incompatible actions $\AA=\{\tA_j\}$ and $\AB=\{\tB_i\}$, the
following mixtures are the same state
\begin{equation}
\sum_j p_j\omega_j=\sum_ip_i'\omega_i'\equiv\omega,
\end{equation}
where
\begin{equation}
\begin{split}
\omega_j&=\frac{\omega(\cdot,\tA_j)}{\omega(\tI,\tA_j)},\quad p_j=\omega(\tI,\tA_j),\\
\omega_i'&=\frac{\omega(\cdot,\tB_i)}{\omega(\tI,\tB_i)},\quad p_i'=\omega(\tI,\tB_i),\\
\omega&\doteq\omega(\cdot,\tI).
\end{split}
\end{equation}
\end{corollary}
\bigskip
Consider now a couple of independent physical systems, say $1$ and $2$.
As we have seen  in Eq. (\ref{nonlocalp}), a probabilistic transformation $\tA$ that occurred on $2$
generally affects the local state of $1$, which then depends on $\tA$ as follows
\begin{equation}
\Omega_{\tA^{(2)}}(\cdot,\tI)\doteq\frac{\Omega(\cdot,\tA)}{\Omega(\tI,\tA)}=\omega^{(1)}_{\tA^{(2)}}
\label{nonlocalp2}.
\end{equation}
\par\bigskip
Finally it is worth mentioning that it is possible to define a ``maximally entangled state'' for a
two-partite system on purely operational grounds as follows 
\begin{definition}[Maximally entangled state]\label{d:maxentang}
A maximally entangled state for two identical independent systems is a pure state $\Omega$ for which the local
state on each system is maximally chaotic, namely
\begin{equation}
\Omega(\cdot,\tI)=\Omega(\tI,\cdot)=\chao{\Stset}.
\end{equation}
\end{definition}
\section{Faithful state}
\begin{definition}[Dynamically faithful state] 
\index{state!dynamically faithful}\index{dynamically faithful state}
We say that a state $\Phi$ of a composite system is {\em dynamically faithful} for the $n$th component
system when acting on it with a transformation $\tA$ results in an (unnormalized) conditional state
that is in  one-to-one correspondence with the dynamical equivalence class $[\tA]$ of $\tA$, namely the
following map is 1-to-1:  
\glossary{\Idx{dyn}$[\tA]_{dyn}$ & dynamical equivalence class of transformations $\tA$}
\begin{equation}
\tilde\Phi_{\tI,\ldots,\tI,\tA,\tI,\ldots} \leftrightarrow [\tA]_{dyn},
\end{equation}
where in the above equation the transformation $\tA$ acts locally only on the $n$th component system.
\end{definition}
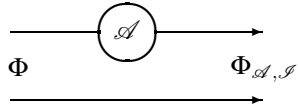
\begin{center}
    \setlength{\unitlength}{800sp}
    \begin{picture}(8745,3219)(931,-3565)
      {\thicklines \put(5401,-1261){\oval(1756,1756)}}
      {\put(1801,-1261){\line(1, 0){2700}}}
      {\put(6301,-1261){\vector(1, 0){3300}}}
      {\put(1801,-3361){\vector(1, 0){7800}}}
      \put(2026,-2611){\makebox(0,0)[b]{$\Phi$}}
      \put(5401,-1486){\makebox(0,0)[b]{$\tA$}}
      \put(9676,-2811){\makebox(0,0)[b]{$\Phi_{\tA,\tI}$}}
    \end{picture}
  \end{center}
\begin{definition}[Informationally faithful state] 
\index{state!informationally faithful}\index{informationally faithful state}
We say that a state $\Phi$ of a composite system is {\em informationally faithful} for the $n$th component
system when acting on it with a transformation $\tA$ results in an (unnormalized) conditional local
state on the remaining systems that is in one-to-one correspondence with the informational equivalence
class $\underline{\tA}$ of $\tA$ (i. e. its propensity), namely the following map is 1-to-1:  
\begin{equation}
\Phi(\cdots,\tA,\cdots)\leftrightarrow \underline{\tA},
\end{equation}
where in the above equation the transformation $\tA$ acts locally only on the $n$th component system.
\end{definition}
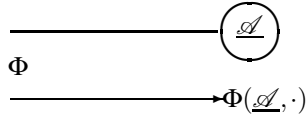
\begin{center}
    \setlength{\unitlength}{800sp}
    \begin{picture}(8745,3219)(931,-3565)
      {\put(1801,-1261){\line(1, 0){6400}}}
      {\put(1801,-3361){\vector(1, 0){6600}}}
      {\thicklines \put(9200,-1261){\oval(1756,1756)}}
      \put(9200,-1486){\makebox(0,0)[b]{$\underline{\tA}$}}
      \put(2026,-2611){\makebox(0,0)[b]{$\Phi$}}
      \put(9676,-3800){\makebox(0,0)[b]{$\Phi(\underline{\tA},\cdot)$}}
    \end{picture}
  \end{center}
\medskip
\par In the following for simplicity we restrict attention to two component systems,
and take the first one for the $n$th. Using the definition \ref{istatecond} of conditional state, we
see that the state $\Phi$ is dynamically faithful when the map $\Phi(\cdot\circ[\tA]_{dyn},\tI)$ is 
invertible over the set of dynamical equivalence classes of transformations, namely when
\begin{equation}
\forall\tA,\;\Phi(\tB_1\circ\tA,\tI)=\Phi(\tB_2\circ\tA,\tI)\quad\Longleftrightarrow\quad\tB_1\in[\tB_2]_{dyn}.
\end{equation}
On the other hand, one can see that the state $\Phi$ is informationally faithful when the map
$\Phi(\underline{\tA},\cdot)$ is invertible over the set of informationally equivalence classes of
transformations, namely when   
\begin{equation}
\forall\tA,\;\Phi(\tB_1,\tA)=\Phi(\tB_2,\tA)\quad\Longleftrightarrow\quad\tB_1\in\underline{\tB_2}.
\end{equation}
\begin{definition}[Preparationally faithful state]\label{prepfaith}
  \index{state!preparationally faithful}\index{preparationally faithful state} We will call a state
  $\Phi$ of a bipartite system {\em preparationally faithful} if all states of one component can be
  achieved by a suitable local transformation of the other, namely for every state $\omega$ of the
  first party there exists a local transformation $\tT_\omega$ of the other party for which the conditioned
  local state coincides with $\omega$, namely
\begin{equation}
\forall\omega\in\Stset\qquad\exists \tT_\omega:\qquad
\frac{\Phi(\tT_\omega,\cdot)}{\Phi(\tT_\omega,\tI)}\equiv\omega.
\end{equation}
\end{definition}
\section{In search for an operational axiom}
In the following I list some possible candidates of operational axioms from which to derive the
quantum superposition principle, namely from which we should be able to determine if a convex set of
states is quantum. We will call a convex set of states $\Stset$ {\em complete quantum convex of
 states} (CQCS or complete QCS) when it coincides with a complete convex set of quantum states on a
given Hilbert space. For example, the Bloch sphere is a CQCS, whereas the unit disk is a QCS, but
not a CQCS. For $n>3$ the $n$-dimensional hypersphere is not a QCS. Similarly, a tetrahedron is
a QCS, but not a CQCS. Notice that the metric is relevant, i. e. an ellipsoid is not equivalent
to the Bloch  sphere, since the antipodal states do not have fixed unit distance.
\par Clearly deriving completeness in terms of an ``operational consistency'' is the difficult part
of the problem, and indeed assuming a kind of completeness for transformations could be just a
restatement of the superposition principle. Following Hardy\cite{Hardyaxioms} we could at most
assume that (a) for any state $\omega\in\Stset$ of a CQCS $\Stset$, the convex set
$\Stset_\omega^\perp$ is a CQCS too, and (b) all pure states in $\Stset$ are connected by an isometric
indecomposable transformation, and these form a continuous group. This, however, leaves out the main
problem of deriving the tensor product structure for independent systems. One would be tempted to
use the following easy axiom
\begin{conjecture}[Existence of maximally entangled states] A convex set of bipartite states
  $\Stset^{\times 2}$ is a QCS if there exist maximally entangled states according to Definition
  \ref{d:maxentang}.  
\end{conjecture}
However, this is not of a truly operational nature. An operational axiom could be a calibrability
axiom of the kind
\begin{conjecture}[Dynamic calibrability] For any bipartite system there exists a pure joint state
  that is dynamically faithful for one of the two systems.
\end{conjecture}
We also conjecture that as a consequence such state is also informationally faithful and
preparationally faithful, or else
\begin{conjecture}[Informational calibrability] For any bipartite system there exists a pure joint state
  that is informationally faithful for one of the two systems.
\end{conjecture}
On the other hand, a preparability axiom could be
\begin{conjecture}[Preparability] For any bipartite system there exists a pure joint state that is
  preparationally faithful for one of the two systems.
\end{conjecture}
\medskip
From one the above calibrability/preparability conjectures the aim would be to prove something as follows
\begin{conjecture}[Dimensionality of composite systems] The informational dimensionality of a
  composite system is the product of their informational dimensionalities. 
\end{conjecture}
This should follow via the equivalence of the dimensionality of the convex cone of
transformations/propensities and that of unnormalized states.  
\par\medskip
Another assertion that is certainly true in the quantum case is 
\begin{conjecture}[Informationally complete discriminating observables] On any bipartite system
  there exists a discriminating observable that is informationally complete for one of the
  components for almost all preparations of the other component.
\end{conjecture}
The above discriminating observable are just the so-called {\em Bell measurements}. Another candidate for
an operational axiom could  be the possibility of achieving teleportation of states
\begin{conjecture}[Teleportation]\label{conj:tele} There exist a joint bipartite state $\Phi$, a joint bipartite
  (discriminating) observable $\AL=\{l_j\}$ and a set of deterministic indecomposable
  transformations $\{\tU_j\}$ by which one can teleport all states as follows 
\begin{equation}
\frac{\omega^{(1)}\Phi^{(2,3)}(l_j^{(1,2)},\cdot\tU_j^{(3)})}{
\omega^{(1)}\Phi^{(2,3)}(l_j^{(1,2)},\tI)}=\omega^{(3)}.
\end{equation}
\end{conjecture}
\begin{conjecture}[Preparability of transformations]\label{conj:teleprep} It is possible to achieve (probabilistically)
  any dynamical equivalence class of transformations using only a fixed action
  $\AA=\{\tA^{(1,2)},\ldots\}$ for a fixed outcome and a fixed partite state $\Phi$, as follows
\end{conjecture}
\begin{equation}
\exists \AA=\{\tA^{(1,2)},\ldots\}:\qquad
\frac{\omega^{(1)}\Phi^{(2,3)}_\tB(\tA^{(1,2)},\cdot)}{
\omega^{(1)}\Phi^{(2,3)}_\tB(\tA^{(1,2)},\tI)}=\omega^{(3)}_\tB.
\end{equation}
As working hypothesis I would like to consider the following combined axioms
\begin{conjecture}[The minimal ``lab'']\label{conj:best} On any bipartite system there exists:
\begin{itemize}
\item [a)] a discriminating observable that is informationally complete for one of the
  components for almost all preparations of the other component.
\item[b)] a pure joint state which, for the same component system in (a) is: dynamically,
  informationally, and preparationally faithful.
\end{itemize}
\end{conjecture}
Another working hypothesis could be that obtained by combining Conjectures \ref{conj:tele} and
\ref{conj:teleprep}, but I think that Conjecture \ref{conj:best} represents the axiom of the most
genuine operational/epistemic nature. 
\begin{theacknowledgments}
Support from the Italian Minister of University and Research (MIUR) is acknowledged under program
Prin 2003. This work has been possible during my last summer visits at Northwestern University,
thanks to the kind ospitality of prof. H. P. Yuen. I thank P. Perinotti, G. Chiribella, C. Fuchs,
K. Svozil, and G. Jaeger for interesting discussions on the first version of the present manuscript,
in particular P. Perinotti, G. Chiribella for a critical analysis, and G. Jaeger for a careful
reading. Definition \ref{def:chaostate} is due to P. Perinotti. 
\end{theacknowledgments}

\newpage
\section*{List of Symbols} 
\begin{center}
\tabletail{&& \\ \hline}
\tablelasttail{&& \\ \hline\hline}
\tablefirsthead{\hline\hline
\glossaryentry{Symbol & Description}{pag}
\hline\hline}
\tablehead{\hline\hline
\glossaryentry{Symbol & Description}{pag}
\hline\hline}
\begin{supertabular}{l|l|r}
\input meta-quantum_proc.glx
\end{supertabular}\end{center}

\end{document}